\documentclass[useAMS,usenatbib,letters] {mn2e}
\usepackage{epsfig}
\usepackage{amssymb}
\usepackage{rotating}
 
\title[Origin of turbulence in the outer Galaxy] {The orientations of molecular clouds in the outer Galaxy: Evidence for the scale of the turbulence driver ?}

\author[Dib et al.] {Sami Dib$^{1}$\thanks{E-mail: sami.dib@cea.fr (SD)}, C. Jakob Walcher$^{2}$, Mark Heyer$^{3}$, Edouard Audit$^{1}$, Laurent Loinard$^{4}$ \\
$^{1}$ Service d'Astrophysique, DSM/Irfu, CEA/Saclay, F-91191 Gif-sur-Yvette, Cedex, France \\
$^{2}$ European Space Agency, Research and Scientific Support Department, Keplerlaan 1, 2200AG, Noordwijk, The Netherlands \\
$^{3}$ Department of Astronomy, University of Massachusetts, Lederle Research Building, Amherst, MA 01003, USA\\
$^{4}$ Centro de Radioastronom\'{i}a y Astrof\'{i}sica, UNAM, Apdo. 72-3 (Xangari), 58089 Morelia, Michoac\'{a}n, Mexico}

\begin{document}
\maketitle

\date{Accepted XXX. Received XXX}

\pagerange{\pageref{firstpage}--\pageref{lastpage}}
\pubyear{2009}
\label{firstpage}

\begin{abstract} 
Supernova explosions inject a considerable amount of energy into the interstellar medium (ISM) in regions with high to moderate star formation rates. In order to assess whether the driving of turbulence by supernovae is also important in the outer Galactic disk, where the star formation rates are lower, we study the spatial distribution of molecular cloud (MC) inclinations with respect to the Galactic plane. The latter contains important information on the nature of the mechanism of energy injection into the ISM. We analyze the spatial correlations between the position angles ($PA$s) of a selected sample of MCs (the largest clouds in the catalogue of the outer Galaxy published by Heyer et al. 2001). Our results show that when the $PA$s of the clouds are all mapped to values into the [$0^{\circ},90^{\circ}$] interval, there is a significant degree of spatial correlation between the $PA$s on spatial scales in the range of 100-800 pc. These scales are of the order of the sizes of individual SN shells in low density environments such as those prevailing in the outer Galaxy and where the metallicity of the ambient gas is of the order of the solar value or smaller. These findings suggest that individual SN explosions, occurring in the outer regions of the Galaxy and in likewise spiral galaxies, albeit at lower rates, continue to play an important role in shaping the structure and dynamics of the ISM in those regions. The SN explosions we postulate here are likely associated with the existence of young stellar clusters in the far outer regions of the Galaxy and the UV emission and low levels of star formation observed with the GALEX satellite in the outer regions of local galaxies.   

\end{abstract} 


\section{INTRODUCTION}\label{intro}
The last few years have witnessed increased efforts in studying the nature and effects of the different energy injection mechanisms into the interstellar medium (ISM) of galaxies. Understanding the nature of the turbulence drivers and their associated energy injection rate has important consequences in improving our understanding of the dissipation of turbulence in the ISM and its impact on star formation. A few recent studies pointed out that kinetic energy might be injected on large scale (kpc scales) into the ISM of dwarf irregular galaxies. Stanimirovic \& Lazarian (2001) analyzed the kinetic energy power spectrum in the Small Magellanic Cloud and did not observe any indication of energy injection up to the largest considered scale ($\sim 4$) kpc. Dib \& Burkert (2004,2005) analyzed the HI gas morphology in Holmberg II (Ho II) and a series of numerical simulations of driven turbulence of the large scale ISM that include, cooling, heating and the gas self-gravity. They performed a comparison between the observations of Ho II and the simulated HI maps by measuring the autocorrelation length on different scales and concluded that turbulence is injected into the ISM of the dwarf irregular, low star forming galaxy Ho II, on a scale of $\sim 6$ kpc. Similar conclusions on the existence of a large scale driving mechanism in the dwarf irregular galaxy DDO 210 have been reached by Begum et al. (2006). Brunt et al. (2009) recently showed that the size-velocity dispersion relations of nearby Galactic molecular clouds (MCs) can only be replicated by models in which turbulence is driven on large scales (i.e., larger the sizes of the clouds themselves). Koda et al. (2006) measured the orientations of MCs in the Galactic Ring region and concluded that the absence of a dominant number of clouds oriented perpendicularly with respect to the Galactic plane is an indication that turbulence in the Galactic ISM must be injected on scales larger than those associated with feedback from massive stars. The latter conclusion is based on the fact that supernova (SN) explosions occurring in the Galactic disk will generate fountain like structures and a large fraction of MCs that are perpendicular to the Galactic plane. However, to date, there is no complete theoretical/numerical model that includes sufficient physics and that possesses enough resolution and that is able to make predictions about the real fractions of  the orientations of MCs in a Milky Way-like Galaxy. In the absence of such information, the conclusions of Koda et al. (2006) still require the quantitative argument that would definitely confirm or exlude stellar feedback as being responsible for the orientations of Galactic MCs.

\begin{figure}
\begin{center}
\epsfig{file=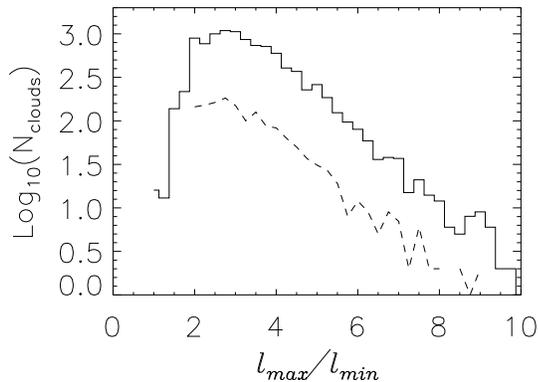,width=\columnwidth}
\caption{Distribution of the axis ratios for of all molecular clouds in the Outer Galaxy molecular cloud catalogue of Heyer et al. (1998, 2001) (full line). $l_{max}$ is the size of the cloud along its major axis and ,$l_{min}$, the size of the cloud along its minor axis. The dashed line corresponds to the sample of clouds with $l_{min} \ge 3$ pc, $l_{max} \ge 6$ pc, and $l_{max}/l_{min} \ge 2$. Note that in the HCS catalogue $l_{max}$ and $l_{min}$ fall in the range [0.01,200.7] pc, and [0.01,43.5] pc, respectively. }
\label{fig1}
\end{center}
\end{figure}

The question of what causes the inclinations of MCs with respect to the galactic plane in the Galaxy and in other galaxies is intimately related to the question of the nature of the dominant energy injection mechanism into the ISM. Numerical simulations and theoretical studies show that feedback from massive stars in the form of stellar winds (Krumholz et al.  2006) or SN explosions (Palou\v{s} et al. 1990, Korpi et al. 1999; de Avillez \& Breitschwerdt 2005; Slyz et al. 2005; Joung \& Mac Low 2006; Dib et al. 2006; Booth \& Theuns 2007; Shetty \& Ostriker 2008) is able to explain the observed velocity dispersions of the HI gas in the inner and intermediate regions of galactic disks and a fraction of the observed velocity dispersion in the outer regions. SN feedback also produces substantial numbers of MCs with non-zero inclinations with respect to the galactic plane. Using a feedback efficiency of 0.25 (each SN injects $E_{SN}=0.25 \times 10^{51}$ erg), Dib et al. (2006) showed that feedback from SNe explosions can maintain a velocity dispersion of $3-5$ km s$^{-1}$ of the HI gas for SN rates such as those prevailing in the outer region of galactic disks; i.e. for SN rates as low as $0.01$ times the Galactic SN type II rate. Dib et al. (2006) attributed the differences between the measured velocity dispersions of the HI gas for such low rates and the observed values ($\sim 5-10$ km s$^{-1}$, Dickey et al. 1990; Dib et al. 2006; Tamburro et al. 2009) in the outer galactic regions to a number of potential effects. The discrepancy can result from the unknown temperatures of the HI gas in the outer regions of galaxies and our ignorance about the true thermal component in the observed line width, an underestimate of the SN efficiency, and/or can be due to the presence of other driving mechanisms in the those regions, such as the magnetorotational instability (Selwood \& Balbus 1999; Kim et al. 2003; Dziourkevitch et al. 2004; Piontek \& Ostriker 2007), large scale gravitational instabilities, non-axisymmetric perturbations, and cloud-cloud collisions (Wada et al. 2002; Dib \& Burkert 2005; Li et al. 2006; Tasker \& Tan 2008; Agertz et al. 2009), tidal interactions and ram pressure effects (Bureau et al. 2004); collisions of high velocity clouds with the galactic disk (Tenorio-Tagle et al. 1987; Santill\'{a}n et al. 2007; Baek et al. 2008) and galactic spiral shocks (Kim \& Ostriker 2006; Bonnell et al. 2006; Kim et al. 2008; Dobbs et al. 2008). 

In the present paper, we complement the work of Koda et al. (2006) by analyzing the position angles of MCs (i.e., their inclinations with respect to the Galactic plane, $PA$) in the outer Galactic disk using the Heyer, Carpenter \& Snell molecular cloud catalogue (HCS, Heyer et al. 2001). As a quantitative test to measure the signature of the turbulence driver(s) in the Outer Galaxy, we measure the spatial correlations between MCs with the same PAs with respect to the Galactic plane. In particular, supernova feedback will generate a network of MCs of various inclinations. If the $PA$s of the clouds are mirrored to values between $0^{\circ}$ and $90^{\circ}$, the spatial scales on which they might be correlated can be compared to the expected sizes of SN remnants in the outer Galactic disk. 

Another possible approach would be to use shell-like structures as a direct tracer of feedback from massive stars. Shells observed in the HI 21 cm line are not an automatic tracer of star formation activity. Although some HI shells are generated by feedback from massive stars, Blitz et al. (2007) discussed the fact that whereas Giant Molecular Clouds are found to be well correlated with high density regions in the HI (see also Margulis et al. 1988), the opposite is not true and many bright HI filaments are found without molecular gas. Dib \& Burkert (2005) and other groups have shown that HI shell-like structures that are not associated with stellar feedback can also be generated by large scale turbulence and gas instabilities. In the SMC, Hatzidimitriou et al. (2005) found that 59 of the 509 HI shell they have catalogued show no signs of being associated to any form of star formation or stellar feedback activities. In Holmberg II, 86 percent of the HI holes do not show signs of being associated with stellar feedback (Rhode et al. 1999). This argument does not apply to CO shells, as it is very likely that large and complete CO shells ($\gtrsim 70$ pc are associated with feedback from massive stars. However, CO shells that would form through feedback by massive stars will generally evolve in a turbulent and highly inhomogeneous medium and may lose their spherical symmetry under the action of hydrodynamical instabilities. This would make their recognition as tracers of feedback by massive stars very problematic. More direct evidence for SN activity would be very difficult to observe in the outer Galaxy. Assuming that the SN rate in the outer Galactic regions is somewhere between 1/10 to 1/100 of the Galactic value, this would result in an SN type II frequency of 1 SN per 500-5000 years since the galactic frequency is 1/(57 years) (Capellaro et al. 1999). It might also be difficult to associate any hot gas with past and present local SN explosions as the hot gas will be easily lost to the halo. One possibility that remains is to to detect remnant stellar clusters associated with the shells. In view of the difficulties in the usage of shell-like structure algorithms to trace the effects of massive stars feedback, we resort to the method described above which is to search for spatial correlations in the molecular clouds orientations with respect to the Galactic disk. 

In \S~\ref{data} and \S~\ref{analysis} we describe the MCs sample and the analysis method, respectively. The results are presented in \S~\ref{results} and in \S~\ref{conc}, we conclude. 

\section{MOLECULAR CLOUD SAMPLE}\label{data}

\begin{figure}
\begin{center}
\epsfig{file=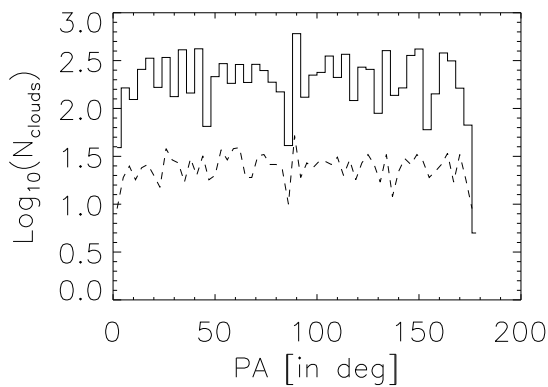,width=\columnwidth} 
\caption{Distribution of the position angles for the entire sample of cores of the HCS catalogue (full line) and the sample of selected clouds with $l_{min} \ge 3$ pc, $l_{max} \ge 6$ pc, and $l_{max}/l_{min} \ge 2$ (dashed line).}
\label{fig2}
\end{center}
\end{figure}

\begin{figure}
\begin{center}
\epsfig{file=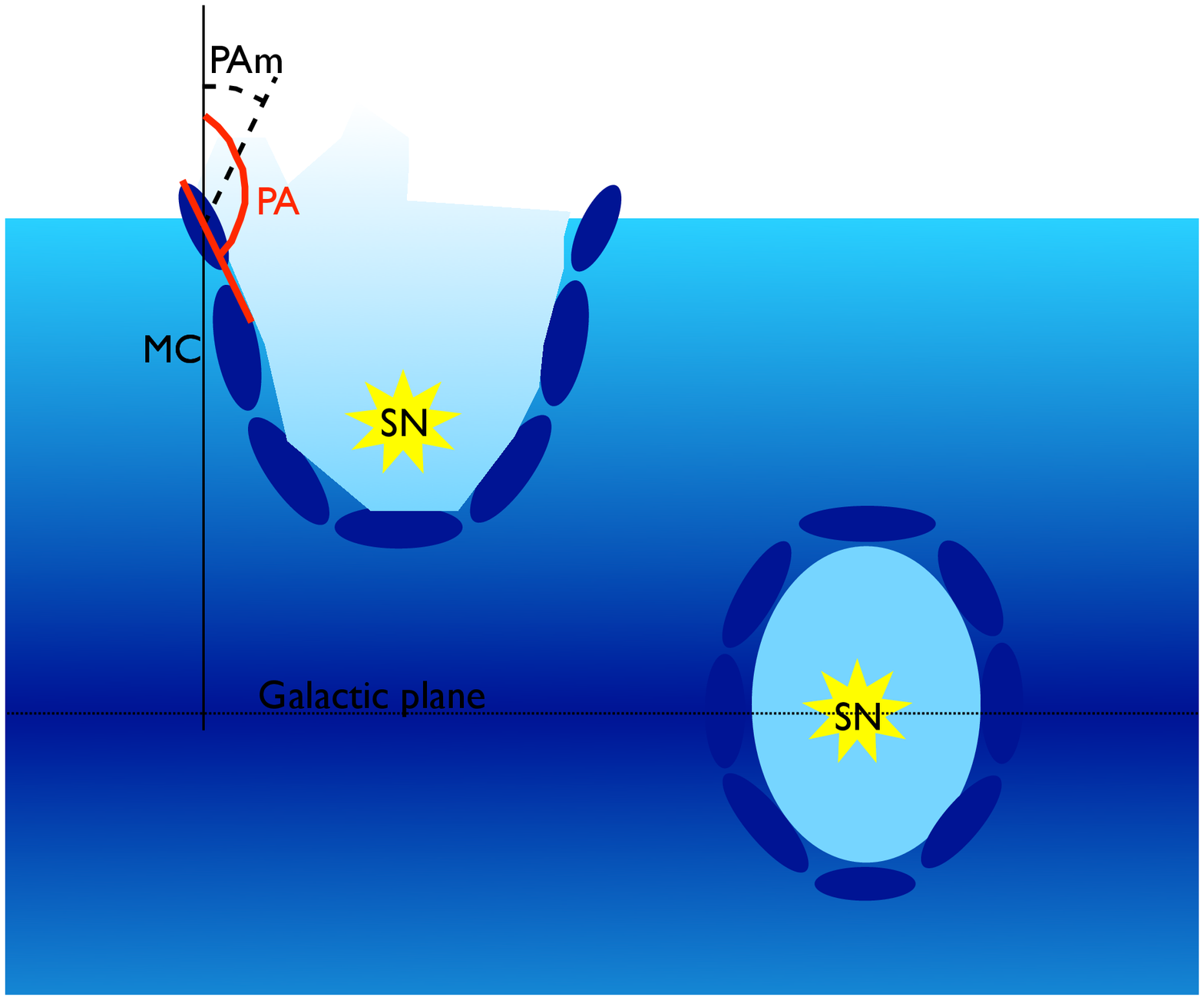, width= 0.85 \columnwidth,angle=270} 
\caption{A schematic, simplified, view of supernova explosions occurring in the Galactic disk  and which illustrates the various possible position angles of molecular clouds generated by the supernova explosion shock. A position angle, PA $ > ~90^{\circ}$ and its mirrored value, PAm, are highlighted.}
\label{fig3}
\end{center}
\end{figure}

For our study of the orientations of MCs in the outer Galaxy, we use the HCS catalogue (Heyer et al. 1998,2001) which is based on the FCRAO $^{12}$CO J=1-0 line survey in the outer Galaxy and contains a total of 10156 objects. The survey covers 336 deg$^{2}$ (40$^{\circ} \times 8.4^{\circ}$) in the second quadrant of the Galaxy with an angular resolution of $45 \arcsec$. An advantage of studying MCs in the outer regions of the Galaxy is that their distances can be uniquely estimated from the observed radial velocity and an assumed rotation curve of the Galaxy (in this case a flat rotation curve in the outer Galaxy with a circular velocity of $v_{circ}=220$ km s$^{-1}$). Individual MCs were identified in this survey as being closed topological surfaces in the $l-b-V_{LSR}$ space at a given threshold of the antenna temperature (1.4 K), where $l$ and $b$ are the Galactic longitude and latitude, respectively, and $V_{LSR}$ is the velocity in the local standard of rest. The HCS catalogue lists a large number of the MCs properties among which are the values of the major and minor axis lengths, $l_{max}$ and $l_{min}$, respectively, the $PA$ which is defined as being the angle between the major axis with respect to the positive Galactic latitude axis measured clockwise, the kinematical distance $d$, and the Galactic coordinates. 

The distribution of axis ratios is shown in Fig.~\ref{fig1}. Most clouds are found to have an axis ratio $l_{max}/l_{min} \gtrsim 2$ (in physical units $l_{max}$ and $l_{min}$ fall in the range [0.01,200.7] pc, and [0.01,43.5] pc, respectively). As pointed out by Heyer et al. (2001), clouds with a large axis ratio have more reliable determinations of their $PA$s. Therefore, we chose to analyze only clouds which have $l_{max}/l_{min} \ge 2$. A second criterion we apply in selecting clouds is that they be large with $l_{min} \ge 3$ pc and $l_{max} \ge 6$ pc. This is to avoid using small clouds whose inclinations with respect to the Galactic plane might significantly differ from the orientation of larger structures they would be part of if a lower threshold for cloud selection were to be used and/or because their actual inclinations might differ from those of the larger structures from which they may have fragmented. Note that at similar kinematical distances, the axis ratios of smaller clouds will be poorly determined compared to those of larger clouds because of the smaller numbers of pixels used in the calculation of the inertial matrix. Applying these two criteria reduces the total number of analyzed clouds to 1485 (R sample). Fig.~\ref{fig2} displays the distribution of $PA$s for all clouds (full line), and for the clouds of sample R (dashed line). This figure shows that MCs in the outer Galaxy, for both distributions, do not have a dominant $PA$ value. They are rather evenly distributed among all possible values of the $PA$s between $0^{\circ}$ and $180^{\circ}$.   

Shell-like structures can be visually observed in the HCS catalogue. Figure 1 in Heyer et al. (2001) which displays the distributions of all CO clouds selected in the velocity interval -110 to -20 km s$^{-1}$ between $l=106^{\circ}$ and $l=142^{\circ}$ shows the existence of large shells centered at ($l,b$) $\sim$ (138$^{\circ},1^{\circ}$), (134$^{\circ}$,0$^{\circ}$), (133.8$^{\circ}$,1$^{\circ}$),(123$^{\circ}$,-1$^{\circ}$), and (111$^{\circ}$,0$^{\circ}$). These shells seem to extend over 1-2 degrees in longitude. However, it is important to note that the CO shells are incomplete which makes the usage of a direct shell-finding algorithm very problematic. In the HI, there are also shells in the same Galactic region that were catalogued by Ehlerova \& Palous (2005). The latter authors do not show a large field of view of the HI gas distribution, but rather a few selected HI shells. Among them, the shell that is observed at ($l,b$)=($124^{\circ}-131^{\circ}$,$0^{\circ}-1^{\circ}$) in figure A.1 of Ehlerova \& Palous seems to coincide in position with two of the CO shells. Note that the few HI shells that are shown seem to be larger, extending over 4-6 degrees in longitude.

\section{ANALYSIS}\label{analysis}
The assumption behind the analysis performed in this work is that a supernova or multiple supernova explosions will lead to compressions of the surrounding gas that would result in the formation of MCs which possess non-zero inclinations with respect to the Galactic plane. Numerical simulations of supernova or multiple supernova explosions in a stratified galactic disk result in mushroom-like shaped cavities with dense and cold, molecular material at their edges (de Avillez \& Breitschwerdt 2005). In order to test the supernova scenario using the $PA$s, it is necessary to mirror the $PA$s and restrict them to the range [0$^{\circ}$,90$^{\circ}$] such that clouds that lay on both sides of the remnant with respect to the normal direction to the Galactic disk will have their concavity pointing in the same direction. This will allow the detection of any correlations between clouds on the scale of individual supernova remnants or of superbubbles. Hence, we calculate the autocorrelation function (ACF) for two cloud samples: sample R, and sample Rm which contains the same clouds as sample R, but where the $PA$s of the clouds in the range [$90^{\circ}$,$180^{\circ}$] are mirrored such as to be restricted to the range of [$0^{\circ},90^{\circ}$] (e.g., the mirror value of $91^{\circ}$ is $89^{\circ}$, see Fig.~\ref{fig3}). The ACF, on a physical scale $L$, is calculated on the discrete sample of clouds using the usual definition: 

\begin{equation} 
 S_{2} (L)=  \frac{\sum_{i} (PA_{i}(r)-\bar{PA})(PA_{i}(r+L)-\bar{PA})}{\sum_{i} (PA_{i}(r)-\bar{PA})^{2}}, 
\label{eq1}
\end{equation}

\begin{figure}
\begin{center}
\epsfig{file= 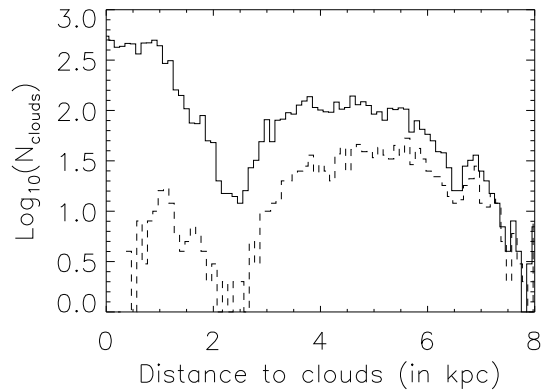,width=\columnwidth} 
\end{center}
\caption{Kinematical distances to the clouds of the entire sample of clouds (full line) in the HCS catalogue (Heyer et al. 1998, 2001) and the sample of selected clouds with $l_{min} \ge 3$ pc, $l_{max} \ge 6$ pc, and $l_{max}/l_{min} \ge 2$ (dashed line).}
\label{fig4}
\end{figure}

\noindent where the index $i$ runs over all MCs in the used sample, $\bar{PA}$ is the average $PA$ in the sample and $L$ is a physical lag. The ACF is evaluated over a series of lags incremented by 100 pc. 
Houlahan \& Scalo (1990) pointed out that, when applied to extended structures, the ACF can be separated into self-terms which are sensitive to their substructure and to to cross-terms  which are more sensitive to their separation. Houlahan \& Scalo (1990) showed that the cross-terms can be overshadowed in a hierarchical-like structure by the self-terms and by other effects such as boundary effects. They point out nevertheless that the usage of the ACF should not be problematic in the case of point like structures, and that in this case the ACF is essentially sensitive to the point-point separation. In our case, since we do not apply the ACF on every single pixel of the clouds (i.e., we do not use their extended gas distribution), we are in the regime of point sources since every cloud is only represented by a unique information which is the $PA$. As we check from the position of each cloud what are the other clouds that are at distances between $r$ and $r+L$ from this cloud, it is necessary to know the cloud-cloud distances. The latter are calculated using the Galactic coordinates and the kinematical distance to each cloud and the cosine law in spherical trigonometry. Fig.~\ref{fig4} displays the kinematical distances to the clouds in the HCS catalogue as derived by Heyer et al. (2001) (full line) and for those of the R (and Rm) samples (dashed line). The cloud-cloud distance, $d_{ij}$, between cloud $i$ and cloud $j$  which are located at distances $D_{i}$ and $D_{j}$ and which have Galactic coordinates ($l_{i},b_{i}$) and ($l_{j},b_{j}$), respectively, is given by:

\begin{eqnarray} 
d_{ij}= [D_{i}^{2}+D_{j}^{2}-2~D_{i} D_{j} cos(b_{i}) cos(b_{j}) cos(l_{i}-l_{j}) \nonumber \\
                                       -2~D_{i} D_{j} sin(b_{i}) sin(b_{j})]^{1/2}.                                       
 \label{eq2}
\end{eqnarray}

\begin{figure}
\begin{center}
\epsfig{file=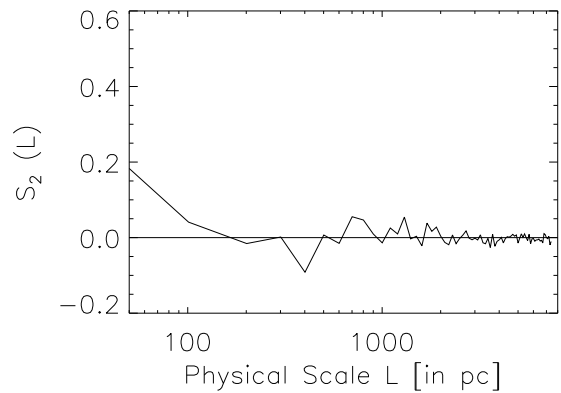,width=\columnwidth} 
\caption{Spatial autocorrelation function of the position angles of the selected clouds in the HCS catalogue (i.e., clouds with $l_{min} \ge 3$ pc, $l_{max} \ge 6$ pc), and $l_{max}/l_{min} \ge 2$.}
\label{fig5}
\end{center}
\end{figure}

\begin{figure}
\begin{center}
\epsfig{file=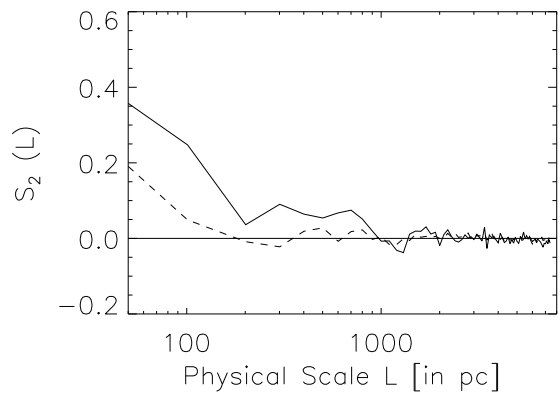,width=\columnwidth} 
\end{center}
\caption{Spatial autocorrelation function (ACF) of the position angles mirrored into the range $[0,90]^{\circ}$ for the selected clouds in the HCS catalogue (clouds with $l_{min} \ge 3$ pc, $l_{max} \ge 6$ pc, and $l_{max}/l_{min} \ge 2$. The dashed line corresponds to the ACF for the same population of clouds but with random position angles (built from the average of 10 individual ACFs with 10 random seed numbers).}
\label{fig6}
\end{figure}

\section{RESULTS}\label{results}
Fig.~\ref{fig5} displays the spatial ACF of the $PA$s for the clouds of sample R. The figure clearly shows that there is no substantial correlation between the orientations of the $PA$s for the clouds of sample R on any physical scale. On the other hand, Fig.~\ref{fig6} displays the spatial correlations of the $PA$s for the Rm sample. In contrast to the un-mirrored case, the mirrored $PA$s for the selected clouds show non-negligible correlations on spatial scales of the order of $\sim 100-800$ pc. We have also checked for the consistency of the results by taking different permutations of $l_{min}$ and $l_{max}$ and of their imposed ratio; i.e., values of ($l_{min}$ (pc),$l_{max}$ (pc))=(2,5), (3,5), (3,7) with and without imposing that $l_{max}/l_{min}\ge 2$, and found no significant variations in the results for Fig.~\ref{fig5} and Fig.~\ref{fig6}. These spatial scales of a few hundred parsecs are well matched by the sizes of supernova remnants in low density environments. Indeed, using analytical calculations, McCray \& Kafatos (1987) estimated  the cooling radius of a supernova shell/supershell to be given by:

\begin{equation}
R_{c}=50 \zeta^{-0.9} (N_{\star} E_{51}^{0.4} n_{0}^{-0.6}) pc,
\label{eq3} 
\end{equation}
 
\noindent where $\zeta$ is the local value of the metallicity, $N_{\star}$ is the number of stars in the stellar cluster with masses greater than $7$ M$_{\odot}$, $E_{51}$ is the energy input of one supernova in units of $10^{51}$ erg, and $n_{0}$, in cm$^{-3}$, is the ambient gas density. The cooling radius marks the end of the adiabatic phase of the expanding shell/supershell and the time at which cooling becomes important in the hot interior of the remnant. Note that the shell/supershell continues to expand afterwards according to the zero-pressure snowplow law with its radius slowly increasing following $R(t)=R_{c} (t/t_{c})^{1/4}$, where $t_{c}$ is the time at which $R_{c}$ has been reached. For an ambient gas density $n_{0} =0.1$ cm$^{-3}$, $N_{\star}=1$, and a solar metallicity, the value of $R_{c}$ is of the order of $\sim 200$ pc, for $N_{\star}=1$, $\zeta=1$, and $n_{0}=0.03$ cm$^{-3}$, $R_{c} \sim 410$ pc, whereas for $n_{0}=0.03$ cm$^{-3}$, N$_{\star}=1$, and a lower metallicity value of $\zeta=0.5$, the cooling radius is $R_{c} \sim 765$ pc. In order to compare the correlations to the noise, Fig.~\ref{fig6} also displays (dashed line) the ACF for the same sample of clouds but with their $PA$s drawn from random values (averaged over 10 iterations). As expected in this case, the ACF tends to zero values on all spatial scales.   

The similarity between the latter physical scales and the physical correlation scales observed in Fig.~\ref{fig6} suggests that supernova driving might be playing a considerable role in shaping the ISM structure and dynamics even in regions of low star formation activity. This is consistent with the results of Dib et al. (2006) who showed that supernova feedback is capable of maintaining turbulent velocity dispersions for the HI gas of the order of $3-5$ km s$^{-1}$ even in regions where the star formation rate per unit area is as low as $10^{-4}$ M$_{\odot}$ yr$^{-1}$ kpc$^{-2}$, corresponding to one hundredth of the Galactic SN rate. The existence of spatial correlations for the sample of mirrored $PA$s which we associate to SN shells and the absence of such correlations for sample R indicates that there are no correlations between the sites of star formation and that gravitational instability occurs in the Galaxy wherever local conditions are favorable. 

\begin{figure*}
\begin{center}
\epsfig{file=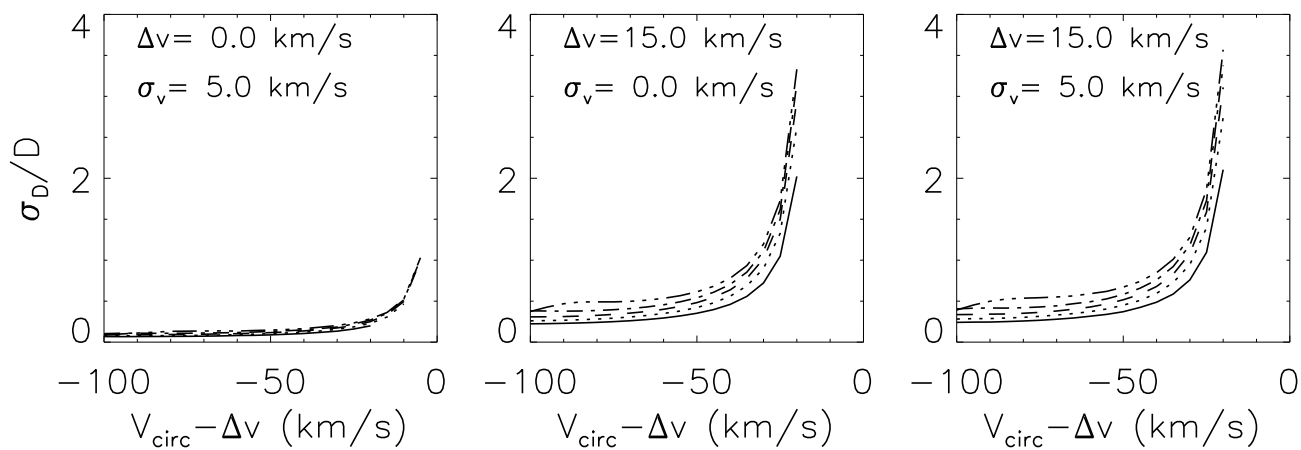,width=\textwidth} 
\end{center}
\caption{Fractional distance errors for clouds in the outer Galaxy (second quadrant) as a function of their velocity and longitude assuming that (left) clouds have a random velocity component $\sigma_{v}=5$ km s$^{-1}$ and no streaming motions $\Delta v=0$ around the flat circular velocity $v_{circ}=220$ km s$^{-1}$, (middle) clouds have streaming motions $\Delta v=15$ km s$^{-1}$ and no random motions ($\sigma_{v}=0$) around the flat circular velocity $v_{circ}=220$ km s$^{-1}$, and (right) clouds have a random velocity component $\sigma_{v}=5$ km s$^{-1}$ and streaming motions with $\Delta v=15$ km s$^{-1}$ around the flat circular velocity $v_{circ}=220$ km s$^{-1}$. The solid, dotted, dashed, dot-dash and triple dot-dash lines correspond to the longitude values of $l=100^{\circ}, 110^{\circ}, 120^{\circ}, 130^{\circ}, 140^{\circ}$, respectively. Note that in the existence of streaming gas motions (middle and right figures), the kinematic distance corresponds always to an overestimate.}
\label{fig7}
\end{figure*}

Note that, in calculating the ACFs, we have made use of the kinematical distances to each of the clouds published by Heyer et al. (2001). The kinematic distance in the outer Galaxy is  an overestimate (e.g., Brand \& Blitz 1993) of the true distance measured by trigonometric parallax measurements (e.g., Hachisuka et al. 2006,2009). For example, more accurate distance estimates using parallax measurements for the W3 region yield a distance of $2.1$ kpc (Hachisuka et al. 2006) which is smaller than the value of $\sim 3.5$ kpc we have used for clouds in that region. The overestimate of the distance by the kinematic distance in the outer Galaxy will depend on the position of the cloud and on the existence of random velocity components and/or velocity offsets at this location with respect to a given Galactic rotation curve. We have calculated  simple models to quantify the kinematic distance uncertainties when using the flat rotation curve (with a circular velocity of $v_{circ}=220$ km s$^{-1}$ similar to the value adopted in Heyer et al. 2001) for objects within the longitude range studied in this paper. The method proceeds as follow: We first calculate a distance assuming pure circular motions for a range of expected velocities and longitudes in the ranges [-5,-100] km s$^{-1}$ and  [$100^{\circ},140^{\circ}$], respectively. In a second step, we add a velocity offset ($\Delta v$) and/or a random component with a dispersion of $\sigma_{v}$ and recompute the distance. The procedure is repeated 8192 times in order to sample the random component of each longitude and velocity. With that, we can derive the root mean square, $\sigma_{D}$, of  the newly calculated distance. Fig.\ref{fig7} shows the fractional distance error $\sigma_{D}/D$ for three sets of $\sigma_{v}$ and $\Delta_{v}$ as a function of the velocity for various values of the longitude $l$. If there are no streaming motions ($\Delta_{v}=0$) but only a random velocity component of dispersion $\sigma_{v}=5$ km s$^{-1}$ (left plot in Fig.\ref{fig7}), the differences between the kinematic distance and the true distance are of the order of $\lesssim 10$ percent for most velocities and get larger for nearby clouds. Such errors could be propagated into the calculation of the inter-cloud distances using Eq.\ref{eq2} on a two-by-two basis. However, the overall uncertainty on the ACF is not expected to be very large and might stretch, for a given physical scale, the x-axis in Fig.\ref{fig5} and Fig.\ref{fig6} by a factor of $10$ percent in one direction or the other. Streaming motions introduce larger but systematic uncertainties which in fact are overestimates of the true distance (models with $\Delta v=15$ km s$^{-1}$, Fig. \ref{fig7} middle and right). The systematic overestimates can be larger than the true distance by factors of up to a few. Since most of the clouds in the sample have velocities $v < -25$ km s$^{-1}$, $\sigma_{D}/D$ is expected to be $\lesssim 1$. To correct for these distance overestimates, the x-axis in Fig.~\ref{fig5} and Fig.~\ref{fig6} should be contracted by a factor of $[(\sigma_{D}/D)+1] \sim 2$. This would cause the mirrored $PA$s to be correlated on scales between $\sim 50-400$ pc. Nevertheless, this would not affect our conclusions.   

In conclusion, our results suggest that SNe do occur in the outer regions of the Milky Way (MW). That stars form at large galacto-centric radii in MW like galaxies is supported by the discovery of young stellar clusters at larger Galactocentric distances (e.g., Carramaza et al. 2008; Yun et al. 2009) and UV observations by the GALEX satellite. The latter show evidence for extended UV emission presumably related to ongoing star formation and feedback from massive stars, at low rates, in those regions (Gil de Paz et al. 2007). This star formation extends beyond the region of the canonical star formation threshold (Martin \& Kennicutt 2001). Observationally, H$_\alpha$ emission, which is indicative of star formation activity is not always associated to the UV emission (Gil de Paz et al. 2007). In fact, Boissier et al. (2007) argue that because of the extremely low star formation rates, the most massive part of the initial mass function is only sparsely sampled and not all star forming complexes host massive stars. SN explosions are at least expected to be associated to a fraction of the star forming sites in the outer galactic regions. Using the Leiden-Dwingeloo HI survey, Ehlerov\'{a} \& Palou\v{s} (2005) analyzed the properties and radial distribution of HI shells in the second quadrant of the outer Milky Way. They found that the radial decrease in the surface density of shells in the outer Galaxy is exponential, with a radial length scale of $\sim 3$ kpc and which is comparable to the length scale of the stellar distribution. The similarity between the length scales of stars and HI shells suggests that at least some shells must be connected to stars and to the effects of stellar feedback. Elmegreen \& Hunter (2006) showed that the radial star formation rates in the disks of spiral galaxies decline gradually with radius rather than being subjected to a sharp cutoff. This is due to the persistence of various gas instabilities in the outer regions such as stellar feedback,  gravitational instabilities, and spiral shocks. 

\section{CONCLUSIONS}\label{conc}
We analyze the spatial correlations of the positions angles ($PA$s) of molecular clouds, i.e., their inclination with respect to the Galactic plane, for a sample of CO clouds located in the outer Galaxy (the CO cloud catalogue of Heyer et al. 1998,2001). When the $PA$s of the clouds that fall in the range [$90^{\circ}$,$180^{\circ}$] are mirrored to values in the range [0$^{\circ}$,90$^{\circ}$], there is a significant degree of spatial correlations between the $PA$s of the largest molecular clouds on spatial scales in the range of $\sim$ 100-800 pc. These physical scales are of the same order as the expected physical sizes of individual supernova (SN) shells in low density environments such as those prevailing in the outer Galaxy and where the metallicity of the ambient gas is of the order of the solar value or smaller. Our results suggest that individual SN explosions continue to occur in the outer regions of the Galaxy, even at lower rates, and may play an important role in shaping the structure and dynamics of the interstellar medium in those regions. 

\section{Acknowledgements}
We thank the referee for constructive and helpful comments and Maheswar Gopinathan for useful discussion. S.D. is supported by the project MAGNET of the ANR and is  grateful for the hospitality of the ESA/RSSD space science faculty during his visit to Noordwijk.  

{}

\label{lastpage}

\end{document}